      [1999/12/01 v1.4c Il Nuovo Cimento]
\documentclass{cimento}

\def\micro      {{\tt micrOMEGAs}}

\title{micrOMEGAs : a tool for dark matter studies}
\author{G.~B\'elanger\from{ins:x}\ETC,
F.~Boudjema\from{ins:x},
A.~Pukhov\from{ins:y},
A.~Semenov\from{ins:z}\\}
\instlist{\inst{ins:x} LAPTH, U. de Savoie, CNRS, B.P.110, F-74941 Annecy-le-Vieux, France 
  \inst{ins:y} Skobeltsyn Inst. of Nuclear Physics, Moscow State Univ., Moscow 119992, Russia
  \inst{ins:z} Joint Inst. of Nuclear Research (JINR) 141980, Dubna, Russia}
\PACSes{\PACSit{12.60,95.35.+d}{\ldots}}
\begin{document}

\maketitle

\begin{abstract}
\micro~ is a tool for cold dark matter (DM) studies  in 
generic extensions of the standard model  with a R-parity like discrete symmetry 
that guarantees the stability of the lightest odd
particle. The code computes the DM relic density, the elastic scattering cross sections of
DM on nuclei relevant for direct detection, 
and the spectra of  $e^+,\bar{p},\gamma$ originating from DM annihilation including porpagation of charged cosmic rays.
The cross sections 
and decay properties of new particles relevant for collider studies are included as well as 
constraints from the flavour sector on  the parameter space of  supersymmetric models. 
\end{abstract}

\section{Introduction}

The existence of a dominant dark matter component in the universe  has been
firmly established by cosmological observations in the last few
years notably by SDSS~\cite{Tegmark:2006az} and
WMAP~\cite{Spergel:2006hy}. Furthermore, the amount of DM today, 
the relic density, has been measured with very good
precision, $\Omega h^2=0.1099\pm 0.0062$~\cite{Dunkley:2008ie}. 
A leading candidate for cold DM is a new weakly interacting stable massive particle 
which naturally provides a reasonable value for the relic
density. It is particulary interesting
that extensions of the standard model (SM)  whose prime goal is to solve the
hierarchy problem can in many cases also provide a viable DM candidate. 
Such particles arise naturally in many
extensions of the standard model ~\cite{Bertone:2004pz} from the
minimal supersymmetric standard
model~\cite{Goldberg:1983nd,Ellis:1983ew} to models of extra
dimensions~\cite{Cheng:2002iz,Agashe:2004ci},
little Higgs models~\cite{Hubisz:2004ft} or models with extended
gauge or Higgs
sectors~\cite{McDonald:1993ex,Barger:2007nv}. In these models
which possess a symmetry like R-parity that guarantees the stability
of the lightest odd particle,
the DM candidate  can be either a Majorana fermion, a Dirac fermion, a vector
boson or a scalar. Their masses range anywhere from a few GeV's to
a few TeV's.

Astroparticle and collider experiments are 
being actively pursued to search for DM. Combined studies of  
DM signals in these different types of experiments should provide
the  necessary ingredients to unravel the nature of DM.
These involve searches for 
DM  either directly through detection of elastic
scattering of the DM with the nuclei in a large detector or
indirectly trough detection of products of DM annihilation
(photons, positrons, neutrinos or antiprotons) in the  Galaxy or
in the Sun. Furthermore colliders are  searching for DM as well as for other new particles 
predicted in  extensions of the standard model.

Sophisticated tools have been developed to perform
computations of DM observables within  R-parity conserving
supersymmetric models, four are publicly available: \micro~\cite{Belanger:2004yn}, {\tt DarkSUSY}~\cite{Gondolo:2004sc},
{\tt IsaTools}~\cite{Baer:2002fv} and {\tt superISO}~\cite{Arbey:2009gu}.  
All  include precise computations of the relic density as well as various constraints on the model from precision
observables or the flavour sector. In addition the first three compute the 
direct and indirect detection rates while \micro~ and \verb|Isatools| provide
collider observables. \micro~ has the added important feature that it can be
easily  generalised to other extensions of the standard model~\cite{Belanger:2006is}. 
This is because 
{{\tt micrOMEGAs}}  is structured around 
{{\tt CalcHEP}}~\cite{Pukhov:2004ca} 
a generic program which once given
a model file containing the list of particles, their masses and
the associated Feynman rules describing their interactions,
computes any cross-section in the model. 
The standard \micro~ routines can then be used to compute the relic
density as well as other DM observables.

\section{Relic density}

A relic density calculation entails solving the evolution equation
for the abundance of DM, $Y(T)$, defined as the
number density divided by the entropy density, ~\cite{Gondolo:1990dk})
\begin{equation}
 \frac{dY}{dT}= \sqrt{\frac{\pi  g_*(T) }{45}} M_p <\sigma v>(Y(T)^2-Y_{eq}(T)^2)
    \label{dydt}
\end{equation}
where $g_{*}$ is an effective number of degree of freedom, $M_p$ is the Planck mass and $Y_{eq}(T)$
the thermal equilibrium abundance. $<\sigma v>$ is the
relativistic thermally averaged annihilation cross-section. The
dependence on the specific particle physics  model  enters only
in this cross-section which includes  all  annihilation and
coannihilation channels,
\begin{equation}
       <\sigma v>=  \frac{ \sum\limits_{i,j}g_i g_j  \int\limits_{(m_i+m_j)^2} ds\sqrt{s}
K_1(\sqrt{s}/T) p_{ij}^2 \sum\limits_{k,l}\sigma_{ij;kl}(s)}
                         {2T\big(\sum\limits_i g_i m_i^2 K_2(m_i/T)\big)^2 }\;,
\label{sigmav}
\end{equation}
where $g_i$ is the number of degree of freedom,  $\sigma_{ij;kl}$
the total cross-section for annihilation of a pair of
odd particles with masses $m_i$, $m_j$ into some
Standard Model particles $(k,l)$, and  $p_{ij}(\sqrt{s})$ is the
momentum (total energy) of the incoming particles in their
center-of-mass frame.

 Integrating Eq.~\ref{dydt} from $T=\infty$ to
$T=T_0$  leads to the present day  abundance $Y(T_0)$
 needed in the estimation of the relic density,
\begin{equation} \label{omegah} \Omega_{LOP} h^2= \frac{8 \pi}{3}
\frac{s(T_0)}{M_p^2 (100{\rm(km/s/Mpc)})^2} M_{LOP}Y(T_0)=
 2.742 \times 10^8 \frac{M_{LOP}}{GeV} Y(T_0)
\end{equation}
where $s(T_0)$ is the entropy density at present time and $h$ the
normalized Hubble constant.

To compute the relic density, \micro~ solves the equation for
the abundance Eq.~\ref{dydt}, numerically without any
approximation. The
computation of all annihilation and coannihilation cross-sections
are done exactly at tree-level. For this we rely  on {{\tt
CalcHEP}}~\cite{Pukhov:2004ca}. 
In the thermally averaged cross-section, Eq.~\ref{sigmav},  coannihilation 
channels are compiled and added only when necessary, that is when the mass difference with the LOP does not exceed a value determined
by the user.

\section{New models}

\micro~ was first developed for the minimal supersymmetric standard model (MSSM). 
In  this model the large number of annihilation channels made automation desirable. 
For this reason \micro~ was based on  {{\tt CalcHEP}}~\cite{Pukhov:2004ca}.  
The generalisation to other 
 particle physics models was then  straigthforward and 
 only requires  specifying the new model
file into {{\tt CalcHEP}}~\cite{Pukhov:2004ca}. 

In order that the program finds the list of processes that need to
be computed for the thermally averaged annihilation
cross-section, relevant for the relic density calculation, one
needs to specify the analogous of R-parity and assign a parity odd
or even to every  particle in the model. The standard model
particles have an even parity.
The lightest odd particle will then be identified with the DM candidate. \micro~ 
automatically generates all  processes of the type
$\chi_i \chi_j \rightarrow X,Y$ where $\chi_i$ designates all
R-parity odd particle and X,Y all R-parity even particles. \micro~ then
looks for s-channel poles as well as for thresholds to adapt the
integration routines for higher accuracies in these specific
regions, and performs the relic density calculation. Note that
the code can also compute the relic density of a charged particle, this quantity 
can for example be used to extract the relic density of the gravitino DM.

 Our approach is  very general and,
as long as one sticks to tree-level masses and cross-sections,
necessitates minimal work from the user beyond the definition of
the model file. However it has been demonstrated that for an
accurate relic density calculation it is necessary in many cases
to take into account higher-order corrections.  In particular
corrections to the mass of either the LOP or of any particle that
can appear in s-channel are important, for example
the corrections to Higgs masses in the MSSM or its extensions.
The large QCD corrections to the Higgs width must also be taken
into account when annihilation occurs near a Higgs resonance. In
general one expects that these  loop corrections can be
implemented via an effective Lagrangian. In practice then it might
be necessary for the user to implement additional routines or
interface other programs to take these effects into account.
This was done explicitly for the NMSSM~\cite{Belanger:2005kh}, the CPVMSSM~\cite{Belanger:2006qa}
and the MSSM with Dirac gaugino masses~\cite{Belanger:2009wf}.

Another advantage of our approach based on a generic program like
{{\tt CalcHEP}} is that one can compute in addition any
cross-section or decay width in the new model considered. In
particular, tree-level cross-sections for $2\rightarrow 2$
processes and decay widths of particles are available.
Furthermore the cross-sections times relative velocity, $\sigma
v$, for DM annihilation at $v\rightarrow 0$ and the yields
for the continuum $\gamma,e^+,\bar{p},\nu$ spectra, relevant for
indirect detection of DM, are also automatically
computed. The procedure to extract the elastic scattering cross section is described next. 


\section{Direct detection}

In direct detection, one measures the recoil energy deposited by
the scattering of DM ($\chi$) with the nuclei. Generically DM-nuclei interactions can
be split into spin independent (scalar) and spin dependent
interations. The scalar interactions add coherently in the nucleus
so heavy nuclei offer the best sensitivity. On the other hand,
spin dependent interactions rely mainly on one unpaired nucleon
and therefore dominate over scalar interactions only for light
nuclei unless scalar interactions are themselves suppressed. In both cases, the cross-section for the DM nuclei
interaction  is typically low, so large detectors are required.
Many experiments involving a variety of nuclei have been set up. 
The best limit for spin independent interactions was reported recently by CDMS  
 with  $\sigma_{\chi p}^{SI}\approx 3.8\times 10^{-8}$ pb for a
DM mass around 70 GeV ~\cite{Ahmed:2009zw}.  This limit already probes a
fraction of the parameter space of DM models.

Many ingredients enter the calculation of the direct detection
rate and cover both astroparticle, particle and nuclear physics
aspects. The detection rate depends on the $\chi$-nucleus cross section which is derived
from  the interaction at the quark level.
The different matrix elements for $\chi q$ interactions  have to be converted into effective
couplings of DMs to protons and nucleons, this is done through coefficients that describe the quark content
in the nucleons. Note that  the  DM have small velocities, therefore  the momentum transfer,
is very small as compared to the masses of the DM and/or
nuclei and the $\chi$-nucleon elastic cross sections can be calculated in
the limit of zero  momentum transfer.  

The distribution of the
number of events over the recoil energy for spin independent interactions reads
\begin{eqnarray}
\label{eq:dNdE:SI}
\frac{dN^{SI}}{dE}=\frac{2M_{det}t}{\pi}\frac{\rho_{0}}{M_\chi}
F_A^2(q)\left(\lambda_p Z + \lambda_n(A-Z) \right)^2I(E)
\end{eqnarray}
where $\rho_0$  is the DM density near the Earth, $M_{det}$ the
mass of the detector, $t$ the exposure time, $I(E)$ is an integral over the velocity distribution
and $F_A(q)$ is the nucleus form factor which  depends on the
momentum transfer. 
 The coefficients $\lambda_p,\lambda_n$ contain all particle physics dependence and are
extracted from the matrix elements for $\chi q$  scattering  coupled with coefficients that describe the quark content
in the nucleons.

Traditionally the
coefficients   $\lambda_{q}$ of the low energy  effective $\chi q$  Lagrangian
are evaluated symbolically using Fiertz identities. Instead in \micro~ 
we use an original approach  that allows to  handle a generic model. 
First we expand the $\chi q$ interactions over a set of basic point-like
operators, only a few operators are necessary in the
$q^2\rightarrow 0$  limit. The same operators also describe the DM-nucleon interactions.
 For example for $SI$ interactions of a Majorana fermion
with quarks or nucleons the effective Lagrangian reads
\begin{eqnarray}
\label{neutralino_si_lgrgn}
  {\cal L}^{SI} &=&\lambda_q \overline{\psi}_\chi\psi_\chi
  \overline{\psi}_q\psi_q
\end{eqnarray}
\micro~ creates automatically a new model file that contains  these operators as new auxiliary vertices in the model.
CalcHEP then generates and calculates symbolically all diagrams for DM - quark/anti-quark
elastic scattering at zero momentum transfer. The interference terms  beetween 
one normal vertex and one auxiliary vertex allow to evaluate numerically the $\lambda_q$
coefficients~\cite{Belanger:2008sj}.  
 Note that in the file that defines the model all quarks should be
defined as massive particles.  Vertices that depend on light quark masses, for example the couplings of Higgs to
 light quarks  cannot be neglected. The dominant term for DM quark scalar
 interactions is proportional to quark masses,  when converting to
   DM nucleon interactions this quark mass will get replaced by a nucleon mass.

\section{Indirect detection}

DM annihilation in the Galactic halo produces pairs of standard model particles.
These particle then hadronize and decay into stable particles that evolve freely in the interstellar medium. 
The final states with  $\gamma$, $e^+$, $\bar{p}$ and  $\bar{D}$ are particularly interesting as  they are the subject of indirect searches.  
Recently  many new results from indirect DM searches have been released notably by 
PAMELA and Fermi~\cite{Adriani:2008zr,Abdo:2009zk}.  
The rate for the production of these particles can be cast into
\begin{equation}
\label{eq:DMflux}
Q({\bf x},E)\;\;=\;\;\frac{1}{2} \langle\sigma v\rangle \left(\frac{\rho({\bf x})}{m_\chi}\right)^2 
\frac{dN}{dE}\;\;,
\end{equation}
where $\langle\sigma v\rangle$ is the  annihilation cross-section 
times the relative velocity of incoming DM particles in the
zero velocity limit.
  $m_\chi$ is the  mass of the DM candidate,  
$\rho({\bf x})$ is the DM density at the location ${\bf x}$ 
and  $dN/dE$ is is the energy distribution of the particle produced in one collision.
The predictions for the energy  spectra depend on non 
perturbative QCD and imply the use of Monte Carlo simulations such as  \verb"PYTHIA".
The annihilation rates are extracted from {\tt CalcHEP} in any model. 
The rate for  photons observed near the center of the galaxy shows a strong dependence on the
halo profile in particular on the DM density near the center, a quantity that still has large uncertainties. 
Different halo profiles are implemented in \micro~. In addition \micro~ provides, in the MSSM, the rate for 
gamma rays that come from direct annihilation of
DM particles. This model-dependent process is  loop-induced and suppressed but 
is nevertheless interesting because of the dramatic signature, a monochromatic gamma-ray line~\cite{Boudjema:2005hb}.

The charged particles generated from DM annihilation propagate through the Galactic halo and their energy spectrum at the Earth differs
from the one produced at the source. 
Charged particles are deflected by the irregularities of the galactic magnetic field, 
suffer energy losses from synchroton radiation and inverse Compton scattering as well 
as diffusive reacceleration in the disk.  Finally
galactic convection wipes away charged particles from the disk. 
Solar modulation can also affect  the low energy part of the spectrum. The  equation that describes the evolution of the 
energy distribution for all particles (protons, anti-protons, positrons) reads
\begin{equation}
\label{eq:propa}
\frac{\partial}{\partial z} \left(V_C\psi\right)
- {\bf \nabla}\cdot\left( K(E) {\bf \nabla} \psi \right)+
\frac{\partial}{\partial E} \left( b(E) \psi  \right) =Q({\bf x},E)
\end{equation}
where $\psi=dn/dE$ is the number density of particles per unit volume and energy.
$Q$ is the production rate of primary particles per unit volume and energy, Eq.~\ref{eq:DMflux}. $b(E)$ is the energy loss rate and 
$K$ is the space diffusion coefficient, assumed homogeneous,
$K(E)=K_0\beta {\cal R}^\delta$ where $\beta$ is the particle velocity and $R$ its rigidity. 
The propagation equation is solved within a semi-analytical two-zone model.
Within this approach the region of diffusion of cosmic rays is represented by a thick disk 
of thickness $2L$ and radius $R=20$~kpc~\cite{Maurin:2001sj} . The thin galactic disk where primary cosmic rays are accelerated lies in the middle 
and has a thickness $2h\approx 200$~pc and radius $R$. The boundary conditions are such that the number density vanishes at $z=\pm L$ and
at $r=\pm R$.
The galactic wind is directed outward along the $z$ direction
so the convective velocity is also vertical and of constant magnitude $V_C(z)=V_C sign(z)$.
The propagation parameters $\delta, K_0,L,V_C$  are constrained by the analysis of the boron to carbon ratio, 
a quantity sensitive to cosmic ray transport~\cite{Maurin:2001sj}. Propagation induces large uncertainties in the prediction of the spectrum from DM annihilation, furthermore
extracting a signal from DM annihilation 
requires a careful understanding of the background as well as of astrophysical sources. Only the DM signal will be
included in the first version of the  indirect detection module to be released soon.

\section{Conclusion}
\micro~ is a comprehensive tool for DM studies in extensions of the SM. The implementation of new models
and new features are being pursued. One objective is a more precise
computation of the relic density. Within the MSSM~\cite{Baro:2007em}
it was shown that the one-loop corrections to the annihilation cross sections could be sizeable as compared with
the precision expected from PLANCK measurements. An interface between the loop-improved computations and
\micro~ is being worked out.

\acknowledgments
This work was  supported in part 
by the GDRI-ACPP of CNRS and by the ANR project {\tt ToolsDMColl}, BLAN07-2-194882.
The work of A.P. was supported by the Russian foundation for Basic Research, grants
RFBR-08-02-00856-a, RFBR-08-02-92499-a and by a State contract No.02.740.11.0244.


\begin{thebibliography}{0}


\bibitem{Tegmark:2006az}
M.~Tegmark {\em et~al.},
\newblock {\em Phys. Rev.}, D74:123507, 2006.

\bibitem{Spergel:2006hy}
D.~N. Spergel {\em et~al.},
\newblock {\em Astrophys. J. Suppl.}, 170:377, 2007.


\bibitem{Dunkley:2008ie}
{\bf WMAP} Collaboration, J.~Dunkley {\em et al.},
 {{\em Astrophys. J.
  Suppl.} {\bf 180} (2009)  306--329}.

\bibitem{Bertone:2004pz}
G.~Bertone, D.~Hooper, and J.~Silk.
\newblock {\em Phys. Rept.}, 405:279--390, 2005.


\bibitem{Goldberg:1983nd}
H.~Goldberg.
\newblock {\em Phys. Rev. Lett.}, 50:1419, 1983.

\bibitem{Ellis:1983ew}
J.~R. Ellis, {\em et al},
\newblock {\em Nucl. Phys.}, {\bf B238}:453--476, 1984.

\bibitem{Cheng:2002iz}
H.-C. Cheng, K.~T. Matchev, and M.~Schmaltz, {\em Phys. Rev.} {\bf
D66} (2002)  036005.
  
  \bibitem{Agashe:2004ci}
K.~Agashe and G.~Servant.
\newblock {\em Phys. Rev. Lett.}, {\bf 93}:231805, 2004.
  
  \bibitem{Hubisz:2004ft}
J.~Hubisz and P.~Meade, {\em Phys. Rev.} {\bf D71} (2005) 035016.
  
 
\bibitem{McDonald:1993ex}
J.~McDonald.
\newblock {\em Phys. Rev.}, D50:3637--3649, 1994.

 \bibitem{Barger:2007nv}
V.~Barger {\em et al.}, {\em Phys. Rev.} {\bf D75} (2007) 115002.

\bibitem{Belanger:2004yn}
G.~B\'elanger, {\it et al.}, {\em Comput.
Phys. Commun.} {\bf 174} (2006) 577.


\bibitem{Gondolo:2004sc}
P.~Gondolo {\em et al.}, {\em JCAP} {\bf 0407} (2004) 008.

\bibitem{Baer:2002fv}
H.~Baer, C.~Balazs, and A.~Belyaev, {\em JHEP} {\bf 03} (2002) 042.

\bibitem{Arbey:2009gu}
  A.~Arbey and F.~Mahmoudi,
  arXiv:0906.0369 [hep-ph].


\bibitem{Belanger:2006is}
  G.~Belanger {\it et al},
  Comput.\ Phys.\ Commun.\  {\bf 176} (2007) 367.

\bibitem{Pukhov:2004ca}
A.~Pukhov, {{\tt hep-ph/0412191}}.

  \bibitem{Gondolo:1990dk}
  P.~Gondolo and G.~Gelmini,
  Nucl.\ Phys.\  B {\bf 360}, 145 (1991).
 

\bibitem{Belanger:2006qa}
   G.~Belanger, F.~Boudjema, S.~Kraml, A.~Pukhov and A.~Semenov,
  Phys.\ Rev.\  D {\bf 73} (2006) 115007.
 
  
  \bibitem{Belanger:2005kh}
  G.~Belanger, F.~Boudjema, C.~Hugonie, A.~Pukhov and A.~Semenov,
  JCAP {\bf 0509} (2005) 001.
  
  \bibitem{Belanger:2008sj}
  G.~Belanger {\it et al.}, 
  Comput.\ Phys.\ Commun.\  {\bf 180} (2009) 747.

\bibitem{Belanger:2009wf}
  G.~Belanger, K.~Benakli, M.~Goodsell, C.~Moura and A.~Pukhov,
  JCAP {\bf 0908} (2009) 027.
  
  \bibitem{Ahmed:2009zw}
  Z.~Ahmed {\it et al.}  [The CDMS-II Collaboration],
  arXiv:0912.3592 [astro-ph.CO].
  
  \bibitem{Adriani:2008zr}
  O.~Adriani {\it et al.}  [PAMELA Collaboration],
  Nature {\bf 458} (2009) 607.
  
  \bibitem{Abdo:2009zk}
  A.~A.~Abdo {\it et al.}  [The Fermi LAT Collaboration],
  arXiv:0905.0025 [astro-ph.HE].

  
  \bibitem{Boudjema:2005hb}
  F.~Boudjema, A.~Semenov and D.~Temes,
  Phys.\ Rev.\  D {\bf 72} (2005) 055024.
  
  
    \bibitem{Maurin:2001sj}
  D.~Maurin, F.~Donato, R.~Taillet and P.~Salati,
  Astrophys.\ J.\  {\bf 555}, 585 (2001).
  
  
  \bibitem{Baro:2007em}
  N.~Baro, F.~Boudjema and A.~Semenov,
  Phys.\ Lett.\  B {\bf 660} (2008) 550.
  
 
\end{thebibliography}
\end{document}